# An Ultra-high Numerical Aperture Metalens at Visible Wavelengths

*Haowen Liang [†], Qiaoling Lin [†], Xiangsheng Xie [‡], Qian Sun [†], Yin Wang [†], Lidan Zhou [†], Lin Liu [†], Xiangyang Yu [†], Jianying Zhou [†], Thomas F Krauss [§], and Juntao Li * [†]*.

[†]State Key Laboratory of Optoelectronic Materials and Technologies, School of Physics, School of Electronics and Information Technology, Sun Yat-Sen University, Guangzhou 510275, China

[‡]Department of Physics, College of Science, Shantou University, Shantou 515063, China

[§]Department of Physics, University of York, York YO10 5DD, UK

* Correspondent author: lijt3@mail.sysu.edu.cn

ABSTRACT: Subwavelength imaging requires the use of high numerical aperture (NA) lenses together with immersion liquids in order to achieve the highest possible resolution. Following exciting recent developments in metasurfaces that have achieved efficient focusing and novel beam-shaping, the race is on to demonstrate ultra-high NA metalenses. The highest NA that has been demonstrated so far is NA=1.1, achieved with a $TiO_2$ metalens and back-immersion. Here, we introduce and demonstrate a metalens with high NA and high transmission in the visible range, based on crystalline silicon (c-Si). The higher refractive index of silicon compared to $TiO_2$ allows us to push the NA further. The design uses the geometric phase approach also known as the Pancharatnam–Berry phase and we determine the arrangement of nano-bricks using a hybrid optimization algorithm (HOA). We demonstrate a metalens with NA = 0.98 in air, a bandwidth (FWHM) of 274 nm and a focusing efficiency of 67% at 532 nm wavelength, which is close to the transmission performance of a $TiO_2$ metalens. Moreover, and uniquely so, our metalens can be front-immersed into immersion oil and achieve an ultra-high NA of 1.48 experimentally and 1.73 theoretically, thereby demonstrating the highest NA of any metalens in the visible regime reported to the best of our knowledge. The fabricating process is fully compatible with CMOS technology and therefore scalable. We envision the front-immersion design to be beneficial for achieving ultra-high NA metalenses as well as immersion metalens doublets, thereby pushing metasurfaces into practical applications such as high resolution, low-cost confocal microscopy and achromatic lenses.

Metasurfaces are artificial sheet materials of sub-wavelength thickness that modulate electromagnetic waves mainly through photonic resonances [1-3]. Their properties are based on the ability to control the phase and/or polarisation of light with subwavelength-scale dielectric or metallic nano-resonators [4, 5]. Correspondingly, metasurfaces are able to alter every aspect of transmitting or reflecting beams, achieving various extraordinary optical phenomena including deflection [6 - 8], retro-reflection [9, 10], polarization conversion [4, 11 - 14], focusing [15 - 17] and beam-shaping [18], with a nanostructured thin film alone. Focusing metasurfaces – namely metalenses – are amongst the most promising optical elements for practical applications [19, 20], e.g. for cell phone camera lenses [21, 22] or ultrathin microscope objectives [23, 24], since their subwavelength nanostructures are able to provide more precise and efficient phase control compared to binary amplitude and phase Fresnel zone plates .

Dielectric materials such as titanium dioxide ($TiO_2$), gallium nitride (GaN) and silicon nitride ($Si_3N_4$) appear to be the best candidates for achieving high focusing efficiency and high numerical aperture (NA) operation, especially at visible wavelengths [22, 25, 26], because of their transparency. However, ultra-high numerical apertures (NA>1) can only be achieved via immersion in high-index liquids [27]. In this case, the refractive index contrast available with dielectric materials is insufficient for achieving high performance. A compromise is to use liquid immersion on the back side of the metalenses, which is known as back-immersion. An example of a back-immersion $TiO_2$ metalens was demonstrated recently, with an NA that is limited to 1.1 at 532 nm wavelength [23]. The design of this lens used the geometric phase, also known as the Pancharatnam-Berry (P-B) phase [28, 29], resulting in an array of "nano-bricks" to control the phase appropriately. In order to achieve higher NA, such metalenses need a smaller period and a correspondingly higher aspect ratio of the nano-bricks to maintain the necessary confinement of the electromagnetic field within the $TiO_2$. The period and aspect ratios are ultimately limited by fabrication constrains [23], which imposes a practical restriction on the maximum NA that can be achieved. By contrast, using a higher refractive index material relaxes these constraints and provides a better solution for maintaining optical confinement in the presence of high-index immersion oil.

Only medium-to-low bandgap semiconductors provide the required high refractive index. Amongst these, amorphous silicon (a-Si) and crystalline silicon (c-Si) are the most promising candidates as both have already been used for the realisation of high performance metalenses in the visible to terahertz frequencies [30 - 32]. The issue with a-Si is that its transmission drops significantly in the visible range due to its high absorption [33, 34]. C-Si, on the other hand, has already demonstrated high efficiency metasurface functionality in the visible regime, with up to 67% diffraction efficiency measured in transmission at 532 nm [35 - 37].

Furthermore, the refractive index of silicon is sufficiently high to realize a front-immersion metalens with liquid immersion. The front-immersion geometry has two advantages: firstly, it enables immersion on the metasurfaces for a doublet structure [6, 38]; secondly, it puts no constraint on the working distance, while a back-immersion metalens requires a very thin carrier to produce the high NA focal spot out with the substrate.

Based on these insights, we now introduce and demonstrate a c-Si metalens with high NA and high transmission in the visible range. By using the geometric phase approached realized with nano-bricks in an arrangement determined by a hybrid optimization algorithm (HOA), we show that a very low loss, high NA metalens can be realized. As an example, we demonstrate an NA = 0.98 metalens in air with a bandwidth (FWHM) of 274 nm and a focusing efficiency of 67% at 532 nm wavelength, which is close to the transmission performance of a TiO$_2$ metalens. Moreover, and uniquely so, our c-Si metalens can be front-immersed into immersion oil and achieve an ultra-high NA of 1.48 experimentally and 1.73 theoretically, thereby demonstrating the highest NA of any metalens in the visible regime reported to the best of our knowledge.

**Results**

***Design of the c-Si based metalens.*** As discussed in ref. [35, 36], the loss of c-Si at wavelengths longer than 500 nm is not negligible only if the geometry of each nano-brick is carefully designed. We then consider that a nano-brick (Figure 1b) can effectively modulate the phase of an incident right-circularly polarized (RCP) beam. Normally, each nano-brick has 5 degrees of freedom: the rotation angle $\theta$, which yields the phase shift according to the P-B phase $\theta(x, y) = \varphi(x, y) / 2$, whereby $\varphi(x, y)$, is the Fresnel phase profile of a metalens, which needs to follow a parabolic profile:

$$\varphi(x, y) = 2\pi - \frac{2\pi n_g}{\lambda}\left(\sqrt{x^2 + y^2 + f^2} - f\right) \tag{1}$$

where $\lambda$ is the target wavelength, $x$ and $y$ are the coordinates of each nano-brick, $n_g$ is the background refractive index around each nano-brick, and $f$ is the designed focal length. Except for the angle $\theta$, the other geometric parameters, i.e. the length $l$, the width $w$, the height $h$, and the center-to-center spacing $a$ of each nano-brick unit cell are determined algorithmically as discussed next.

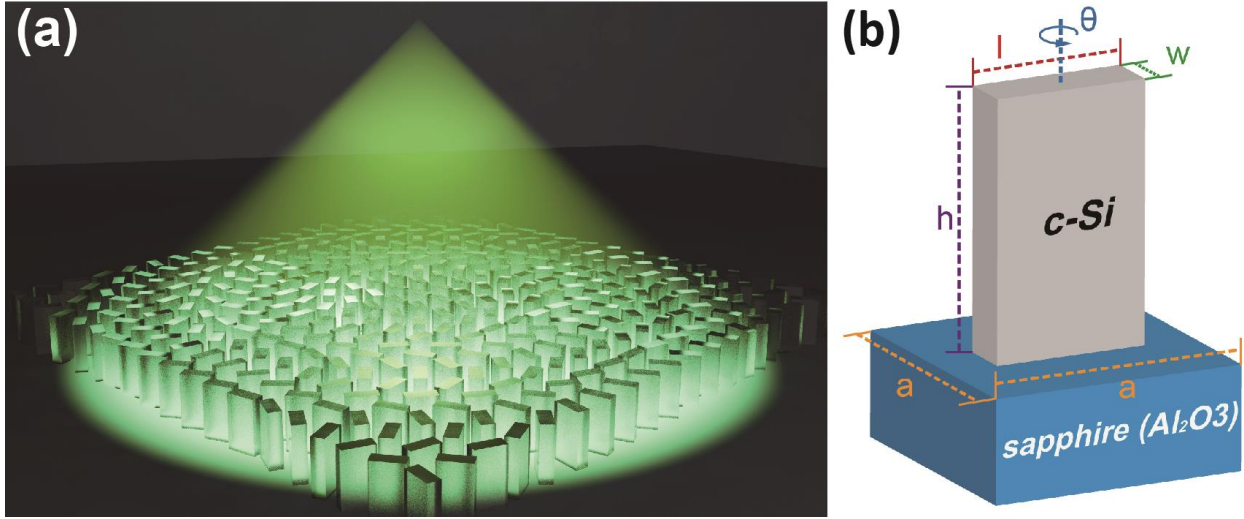

**Figure 1**. (a) Schematic of the c-Si metalens. For the immersion purpose, the immersion liquid is placed on the same side as the nano-bricks (b) The metalens consists of c-Si nano-bricks on a sapphire substrate. The length $l$, width $w$, height $h$ and the center-to-center spacing $a$ of the unit cell are optimized by the hybrid optimization algorithm.

Apart from the phase, the other four degrees of freedom need to be determined algorithmically; the traditional sweep algorithm is time consuming when multiple parameters are involved, however. Other optimization algorithms [39 - 42], including differential evolution (DE), genetic algorithm (GA), particle-swarm optimization (PSO) and adaptive simulated annealing (ASA), can all be utilized. It is not clear, however, whether they will be able to optimize both the phase and the transmission into the global optimum. Here, we conduct a self-adaptive hybrid optimization algorithm (HOA) which is combined with DE, GA, PSO and ASA to efficiently search the global optimized geometry of a nano-brick to avoid large loss of c-Si as well as to determine the exact phase shift. The normalized figure of merit (*FoM*) for single wavelength in the optimization process is therefore defined as:

$$FoM = 2 - e^{|\Phi - \Phi_0|} - \left| \frac{T_{TE} + T_{TM}}{2} - T_0 \right| \Big/ T_0 \qquad (2)$$

where $\Phi$ and $\Phi_0 = 0$ are the calculated and the target phase, respectively; $T_{TE}$, $T_{TM}$ and $T_0 = 1$ are the transmission of the TE mode, the TM mode and the expected total transmission, respectively.

The hybrid optimizing process proceeds as follows: the initial population in DE, GA, PSO and ASA are randomly distributed within an area subjected to so-called Pareto Frontier, which indicates that any further improvement of the solution in terms of one objective (e.g. the phase shift) is likely to be compromised by the degradation of another objective (e.g. the transmission). The fitness of these individuals is weak and dose not yet contain the actual set of non-dominated solutions. Then the HOA begins the first optimizing generation, activating the first optimizer (DE in this work) and pushing the population in DE towards the Pareto Frontier. Once the DE optimization stops at the local extremum, the next optimizer (GA) is activated to continuously push the corresponding population towards the Pareto Frontier; additionally, some "immigrants" in the first optimizer are introduced into the populations that can provide alternative good solutions to the ones already being explored by one of the algorithms. These immigrants can steer the population into other unexplored areas of the search space, avoiding for the algorithm to get trapped in local optima hence increasing the chances of locating the global optimum. The processes repeats in order for PSO and ASA (as shown in Figure 3a to 3d). When the first optimizing generation ends, the HOA compares the most efficient points according to the *FoM*, in the four optimization algorithms to check their correlation. If these efficient points do not obviously point to the global optimum, the HOA begins the next optimizing generation until all optimizers converge on the Pareto Frontier and return the most efficient point with the largest *FoM* (Figure 2e). The geometry of the unit cell can then be obtained by unwrapping the elements in the combination of ($\Phi$ ($l$, $w$, $h$, $a$), $T$ ($l$, $w$, $h$, $a$)) giving by the most efficient solution.

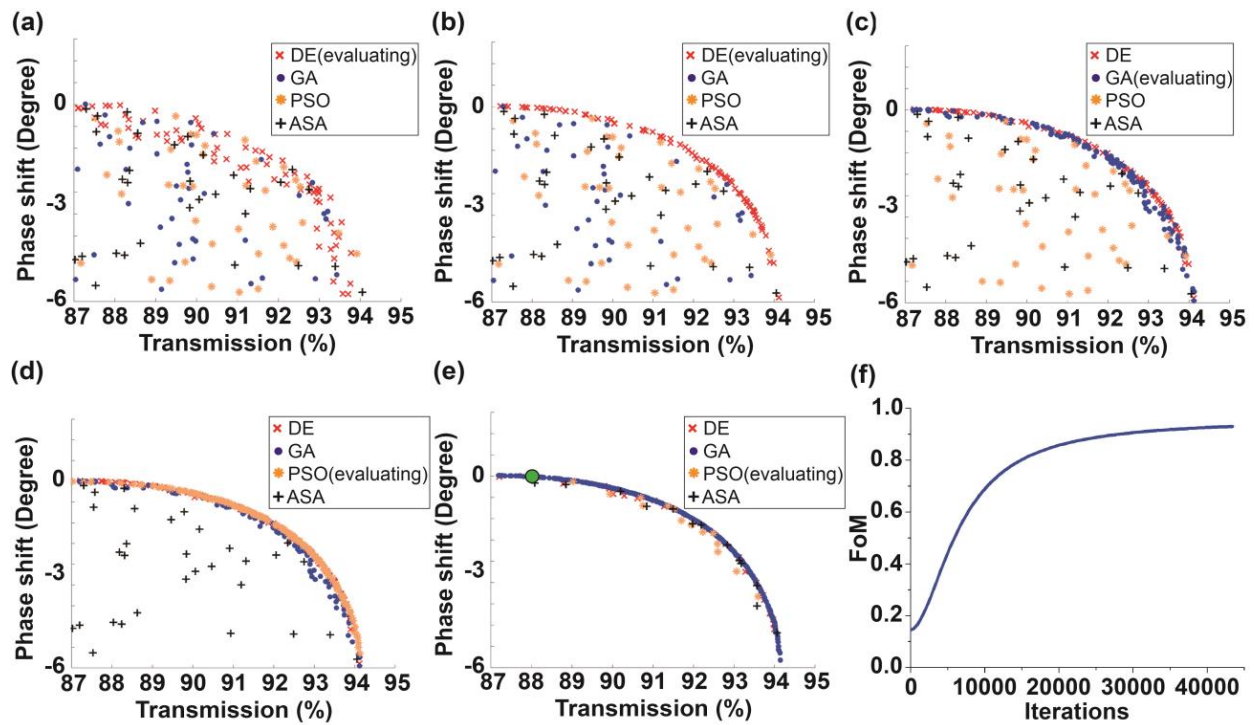

**Figure 2** (a) the initial population in DE, GA, PSO and ASA, showing weak fitness of each individual algorithm. (b) The population distribution in generation 1 with 2434 iterations, where DE is evaluated. (c) The population distribution in generation 1 with 6308 iterations, where GA is evaluated. (d) The population distribution in generation 1 with 12996 iterations, where PSO is evaluated. (e) The HOA converges in generation 3 with 43946 iterations, showing the global best result at ($\Phi$ = 0.002 °, $T$ = 88%). (f) Plot of the *FoM* evolution in terms of the iterations, only showing the best *FoM* of each iteration. The final *FoM* is 0.93, indicating a transmission of 88% for c-Si at the wavelength of 532 nm, which is remarkably high. Over the course of multiple iterations (Figure 2f), the parameters for the metalens designed at $\lambda$ = 532 nm are: $l$ = 160 nm, $w$ = 20 nm, $h$ = 500 nm, and $a$ = 220 nm, respectively. The designed immersion metalenses can then be arranged according to eq. 1 with the corresponding nano-bricks.

***Simulation of the c-Si based metalens.*** The performance of the immersion metalens is evaluated using commercial software (Lumerical Inc.) based on 3D finite difference time domain (FDTD) simulations. Because of the memory space limitation, a small metalens (50 μm × 50 μm) with the same NA as the fabricated one ($\emptyset$ = 1 mm, NA = 0.98) is simulated (shown in Figure 3a to 3c). The full width at half maximum (FWHM) of the focal spot for a wavelength of 532 nm is as small as 277 nm (Figure 3c) in the focal plane. To differentiate between the power transmitted through the lens and the power directed by the lens toward the focus, the focusing efficiency can be defined as the fraction of the incident light that passes through a circular iris in the focal plane with a radius equal to three times the FWHM spot size [30]. Accordingly, the simulation indicates a 71% focusing efficiency for our metalens. Furthermore, the focusing efficiency of the light contributing to NA from 0.975 to 0.98 is 58.4% (The calculation is shown in the Supporting Information). The focusing efficiency for the same lens in immersion oil with $n_g$ =1.512 is also simulated; the efficiency then drops to 56%, while the FWHM of the focal spot reduces to 207 nm (Figure 3f), as expected, leading to an NA = 1.48, which shows that ultrahigh NA metalenses are indeed possible.

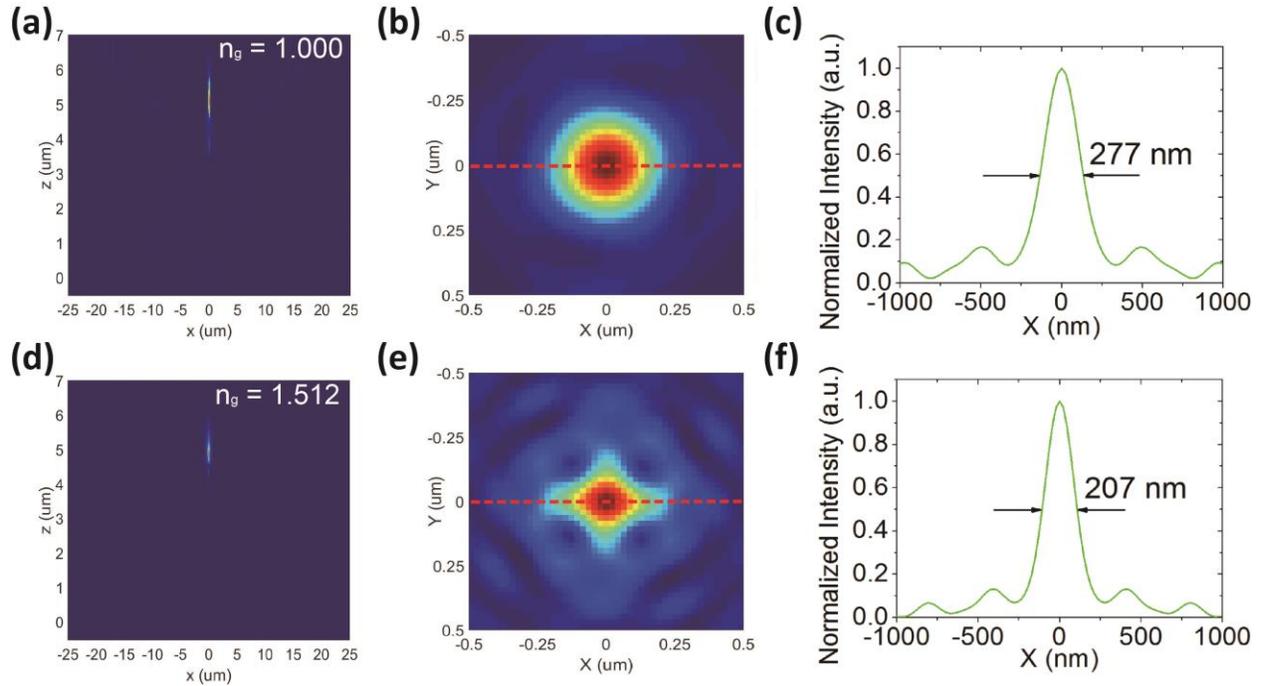

**Figure 3**. Simulated focusing characterization of high-NA metalens with incident RCP light at 532 nm. (a) Focusing performance of the metalens in air. The focal point is at a distance $z$ = 5.1 μm from the lens surface. (b) Normalized intensity profile of the focal spot at the focal plane. (c) Intensity distribution in the focal plane along the dotted red line on (b), showing a 277 nm FWHM of the focal spot. (d) – (f) Corresponding analysis of panels (a) – (c) for a metalens in immersion oil with $n_g$ = 1.512, showing a 207 nm FWHM of the focal spot with NA = 1.48. The clova-leaf shape is due to the fact that a square lens was used in the model. The calculation of NA is shown in the Supporting Information.

***Fabrication and characterization of the metalens.*** To verify the design, a circular high-NA metalens with diameter of 1 mm was fabricated on a crystalline silicon – on – sapphire (SOS) wafer with 500 nm thick c-Si on 500 μm thick sapphire (from Univ. Wafer). The pattern is defined in high resolution negative resist (Hydrogen silsesquioxane, HSQ, 200-nm thick film, Dow Corning) by electron beam lithography (EBL, Raith Vistec EBPG-5000plusES) written at 100keV.

The development is performed in tetramethylammonium hydroxide (TMAH) and the patterned sample is etched in an Inductively Coupled Plasma tool (ICP, Oxford Instruments PlasmaPro System 100ICP180) with HBr chemistry, which has been optimized for vertical, smooth sidewalls and a high aspect ratio of the nano-bricks. Finally, the redundant HSQ is removed by hydrogen fluoride (HF) acid. Scanning electron microscope (SEM, Zeiss Auriga) images of the metalens are shown in Figure 4.

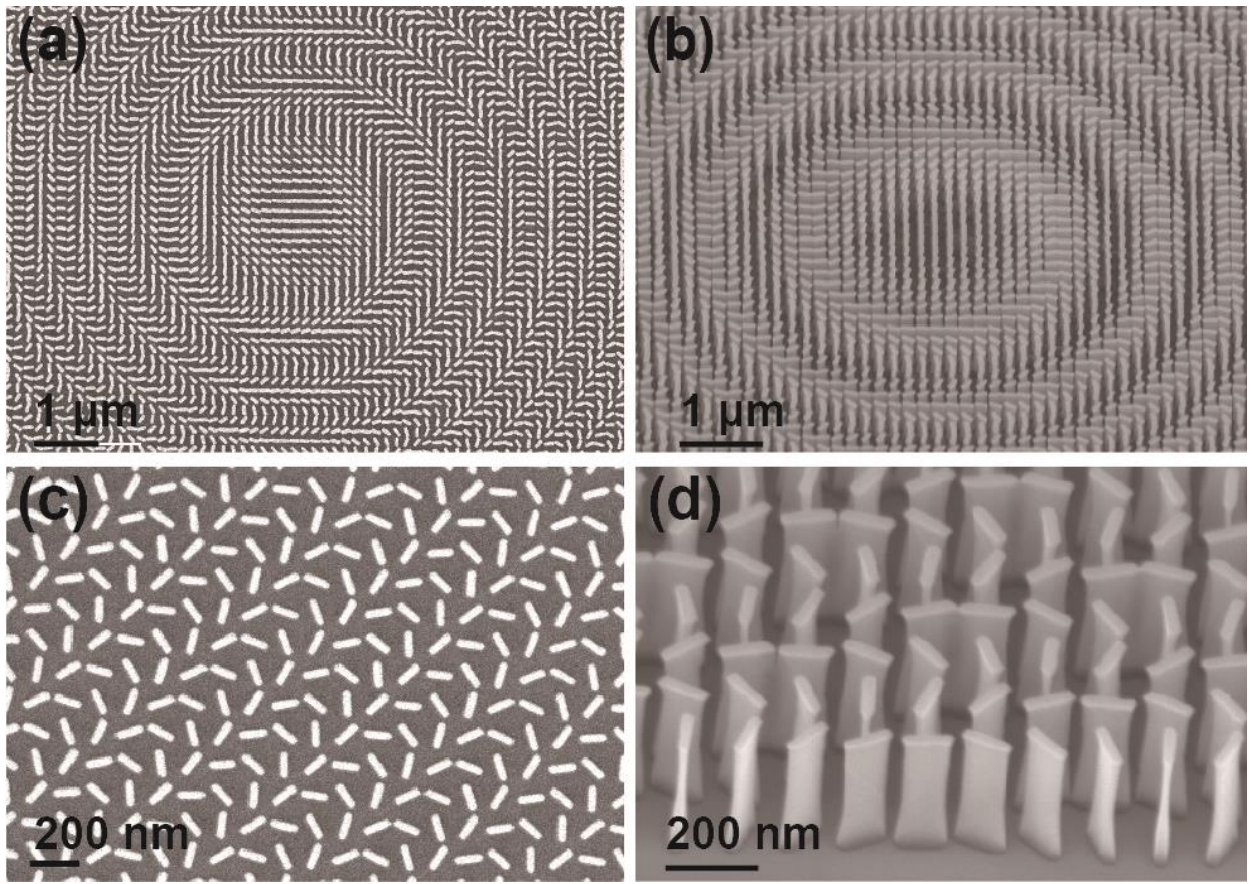

**Figure 4** SEM images of the high-NA metalens. (a) Top-view and (b) 45 °tilted-view of the central section of the metalens. (c) Top-view SEM image of a portion of the metalens at a higher magnification, displaying each individual nano-brick. (d) 45 ° tilted-view of the portion of the metalens at a higher magnification than that in (b), showing the erection and sidewalls of the nano-bricks.

The characterization of the focusing performance is then conducted with a combination of a circular aperture, 100X objectives, a tube lens and a CCD camera at 532 nm wavelength (Figure 5a). The measured focal distance is $f = 105$ μm, corresponding to an NA = 0.98 (see Supporting Information for a detailed calculation). Figure 5b shows the focal spot intensity profile of the metalens in different horizontal planes along the z axis. The measured focusing efficiency is 67%, which is very close to the performance of a TiO$_2$ metalens [22], yet only a c-Si metalens can be operated with front-immersion (see below). The FWHM of the measured focal spot (Fig. 5c) is 274 nm compared to the value of 279 nm expected for a diffraction-limited focusing spot: FWHM = 0.51λ/NA [23, 27]. The simulated focal spot is 277 nm according to the simulation in the above section, so again, the values match very closely.

For the immersion experiment, the same c-Si metalens is front-immersed in oil of refractive index $n_g = 1.512$. This front-immersion metalens exhibits a focusing efficiency of 48% and yields a further reduced focal spot of 211 nm FWHM compared to a value of 182 nm expected for a diffraction-limited spot (Fig. 5d), leading to an ultra-high NA = 1.48 (see Supporting Information for a detailed calculation), which is significantly higher than the best result achieved with a back-immersion metalens in TiO$_2$ (NA = 1.1) [23]. Interestingly, the TiO$_2$ lens has a very similar focusing efficiency of 50% in that case. Our measured NA = 1.48 highlights the very good agreement between theory and experiment and indicates that even higher NAs of 1.73 may be achievable, as indicated by our model (see Supporting Information for a detailed calculation) for the case of using immersion liquids of the same refractive index as the sapphire substrate (1.76).

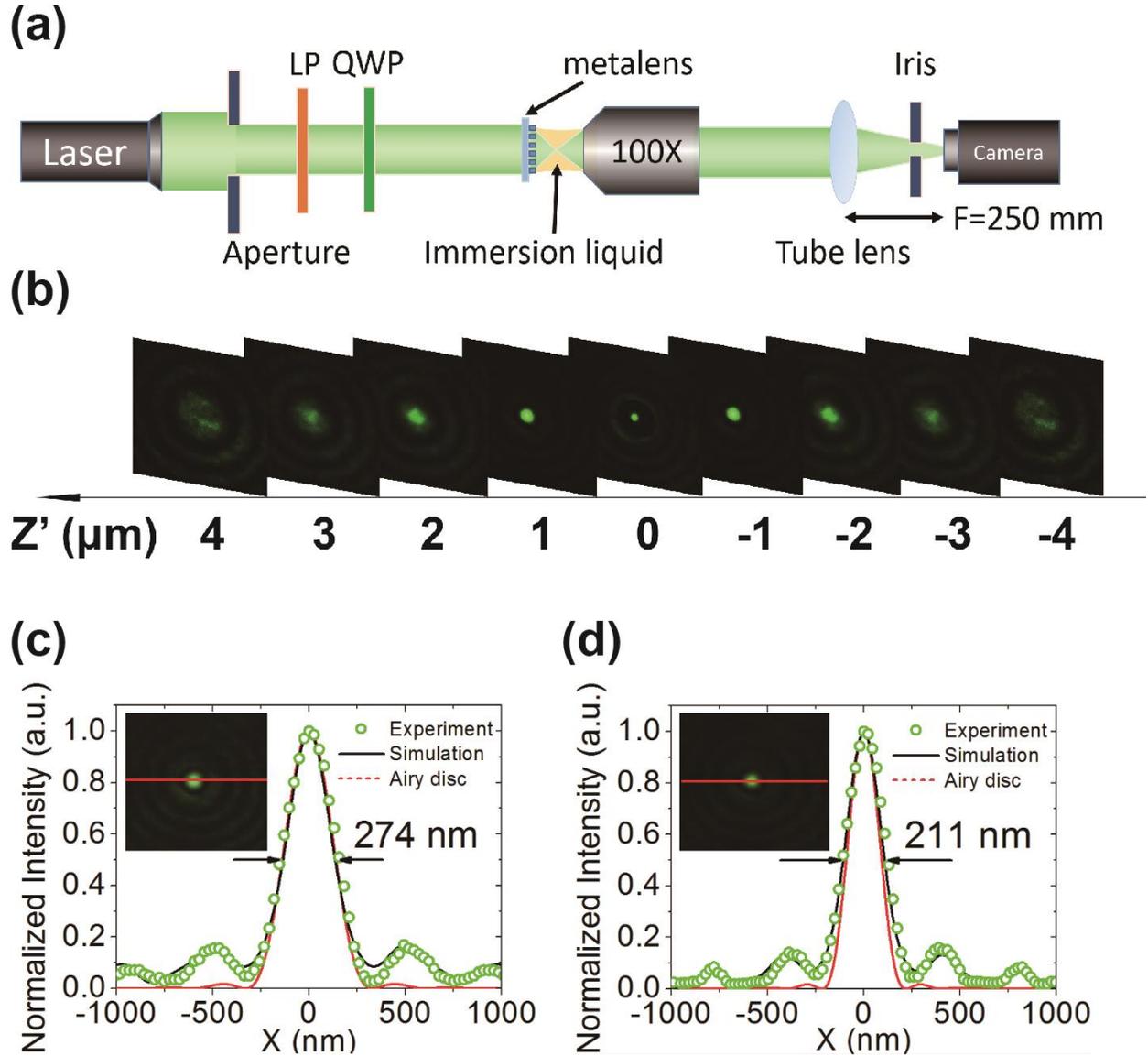

**Figure 5** (a) Experimental setup for measuring the size of the focal spot. The laser beam is collimated and passes through an aperture, a linearly polarizer (LP) and a quarter-waveplate (QWP) to generate RCP light. A 100X immersion objective and a tube lens (L) with focal length of 250 mm are used to collect the transmitted light. An iris with radius equaling to three times of the FWHM spot size is used for proper calibration of the focusing efficiency. (b) Experimentally measured depth of focus of the c-Si metalens showing the evolution of the beam from 4 μm in front of the focal point to 4 μm behind it. (c) Experimental intensity distribution (green dots) of the focal spot in air and (d) in immersion oil with $n_g = 1.512$. Inset: experimental focal spot captured by the CCD.

**Discussion**

Table 1 summarizes the performance parameters of other experimentally reported metalenses and compares them to our result. The key novelty we introduce here is that our metalens is able to operate in immersion liquid of very high refractive index (n=1.512) with a focusing efficiency of 48% and a FWHM spot size of 211 nm, We show that despite the relatively high absorption loss in the visible range, c-Si can be used to design and make high efficiency metalenses. The very high refractive index of c-Si provides a fundamental advantage for metalens design as it affords the use of high refractive index immersion liquids while providing the necessary confinement of light. This unique material characteristic highlights that a c-Si metalens has the potential not only for immersion applications, but also for multi-layered metasurfaces that consists of different dielectric materials. C-Si based metalenses also extend the application-space of metalens doublets to immersion applications, which is of great significance for super-resolution microscopy, achromatic lens [41 - 46], optical trapping or sub-water imaging. Finally, crystalline silicon is readily available and CMOS compatible, so the fabrication of these metalenses can easily be scaled up.

**Table 1.** Summary of previously reported experimental transmission metalenses

| Reference | Materials | Wavelength (nm) | FWHM spot size ($\lambda$) | NA | Efficiency |
|---|---|---|---|---|---|
| Chen et al. [23] | TiO$_2$ | 532 | 0.57 | 1.1 ($n_g$=1.52) | 50 % |
| Khorsaninejad et al. [22] | TiO$_2$ | 405, 532, 660 | 0.67, 0.63, 0.73 | 0.8 (in air) | 86, 73, 66% |
| Paniagua-Domínguez et al. [24] | a-Si | 715 | 0.54 | 0.99 (in air) | 35% |
| Lin et al. [33] | p-Si | 500 | 1.27 | 0.43 (in air) | 38% |
| Wang et al. [46] | GaN | 400 - 660 | ~ 5.3 | 0.106 (in air) | 40% |
| Arbabi et al. [30] | a-Si | 1550 | 0.57 | 0.97 (in air) | 42% |
| This work (in air) | c-Si | 532 | 0.52 | 0.98 (in air) | 67% |
| This work (in $n_g = 1.512$) | c-Si | 532 | 0.40 | 1.48 ($n_g = 1.512$) | 48% |

AUTHOR INFORMATION

**Corresponding Author**

*E-mail: lijt3@mail.sysu.edu.cn

**Author Contributions**

H. L. and J. L. conceived the concept. H. L. performed numerical simulations, structural designs, sample fabrication, experimental characterization and wrote the manuscript. Y. W., L. Z. and L. L fabricated the samples, Q. L., X. X. and Q. S. performed numerical calculations. J. L., T. F. K., J. Z. and X. Y. participated in planning the experiments and supervised the project. All authors participated in discussions and contributed to editing of the manuscript.

**Notes**

The authors declare no competing financial interest.

**ACKNOWLEDGMENT**

This work is supported by National Key R&D Program of China (2016YFA0301300), National Natural Science Foundation of China (11534017, 11704421, 11674402, 11761131001, 91750207), Guangdong Science and Technology Project (201607010044, 201607020023), and Three Big

Constructions—Supercomputing Application Cultivation Projects. TFK acknowledges funding from the EPSRC Program Grant EP/P030017/1 "Resonant and shaped photonics" and the Royal Society Wolfson Research Merit Award scheme.

# Supporting Information for An Ultra-high Numerical Aperture Metalens at Visible Wavelengths

*Haowen Liang [†], Qiaoling Lin [†], Xiangsheng Xie [‡], Qian Sun [†], Yin Wang [†], Lidan Zhou [†], Lin Liu [†], Xiangyang Yu [†], Jianying Zhou [†], Thomas F Krauss [§], and Juntao Li * [†].*

[†]State Key Laboratory of Optoelectronic Materials and Technologies, School of Physics, School of Electronics and Information Technology, Sun Yat-Sen University, Guangzhou 510275, China

[‡]Department of Physics, College of Science, Shantou University, Shantou 515063, China

[§]Department of Physics, University of York, York YO10 5DD, UK

* Correspondent author: lijt3@mail.sysu.edu.cn

## 1. Numerical aperture

The numerical aperture (NA) is determined by the maximum diffraction angle $\theta$ at the edge of the metalens and the refractive index $n_g$ of its surrounding immersion materials:

$$\text{NA} = n_g \sin \theta$$

(S1)

where, $\tan\theta$ is defined by the ratio between radius $r$ and its focal distance $f$ of the metalens. The calculated NA in the manuscript are shown as Table S1

Table S1. Calculated NA in the simulation and experiment

|  | $r$ (μm) | $f$ (μm) | $n_g$ | NA |
|---|---|---|---|---|
| Metalens in air (simulation) | 25 | 5.1 | 1 | 0.98 |
| Metalens in immersion (simulation) | 25 | 4.9 | 1.512 | 1.48 |
| Metalens in air (experiment) | 500 | 105 | 1 | 0.98 |
| Metalens in immersion (experiment) | 500 | 102 | 1.512 | 1.48 |

An effective method to clarify that our metalens enables the focusing with an exact 0.98 NA is blocking the light with low spatial frequencies while letting that with high spatial frequencies through. Hence, an annular phase mask (as shown in the inset of Fig. S1(a), the white color shows that light wave is able to transmit through while the black color is region made of 2 μm thick c-Si on the back of the substrate to absorb the input light is used in the simulation of 50 μm ×50 μm metalens mentioned in this letter. In order to block the low spatial frequencies, the radius ratio of the annular mask, which is defined by the inner radius $r_1$ over the outer radius $r_2$, is 0.9, which means only focus light contributed to NA between 0.975 to 0.98 can pass the metalens. An "optical needle" with focal length at 5.1 μm is obtained (Fig. S1(a)). 8.7% of the input power is focused on the focal spot. Considering 85.1% of the input power is blocked by the annular phase mask, the focusing efficiency of the high NA regime is 58.4%. As shown in Fig. S1(b), the shape of the focal spot is elliptical, as explained in Ref. S1. Furthermore, the focal spot size along the minor axis of the elliptical focal spot is 251 nm, which is smaller than the metalens without annular phase mask [S1].

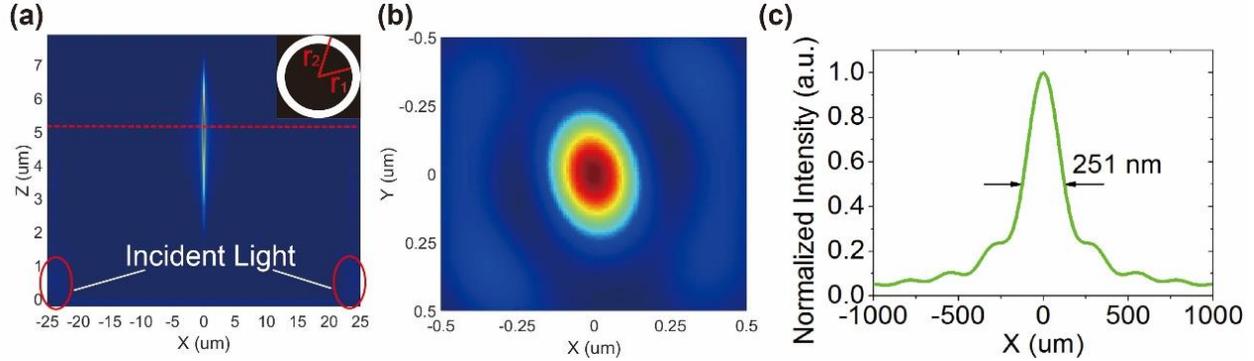

**Figure S1**. (a) The focusing performance of the metalens that only the light with high spatial frequencies can transmit through. The inset shows an annular phase mask with a radius ratio of 0.9 blocking most of the light with low spatial frequencies. (b) Intensity distribution in the focal plane along the dotted red line on (a), and (c) its intensity distribution along the minor axis of the elliptical focal spot, showing a 251 nm FWHM.

## 2. Focal spot size

The focal spot sizes in terms of NA are shown in Figure S2. Here, the simulated, experimental and theoretical focal spot sizes are from the FWHM of the focal spot according to the FDTD simulation in the manuscript, measurement in Figure 5 (c) and (d), and the diffraction-limited calculation of the Airy disc: $0.51\lambda/\text{NA}$ [S2, S3], respectively.

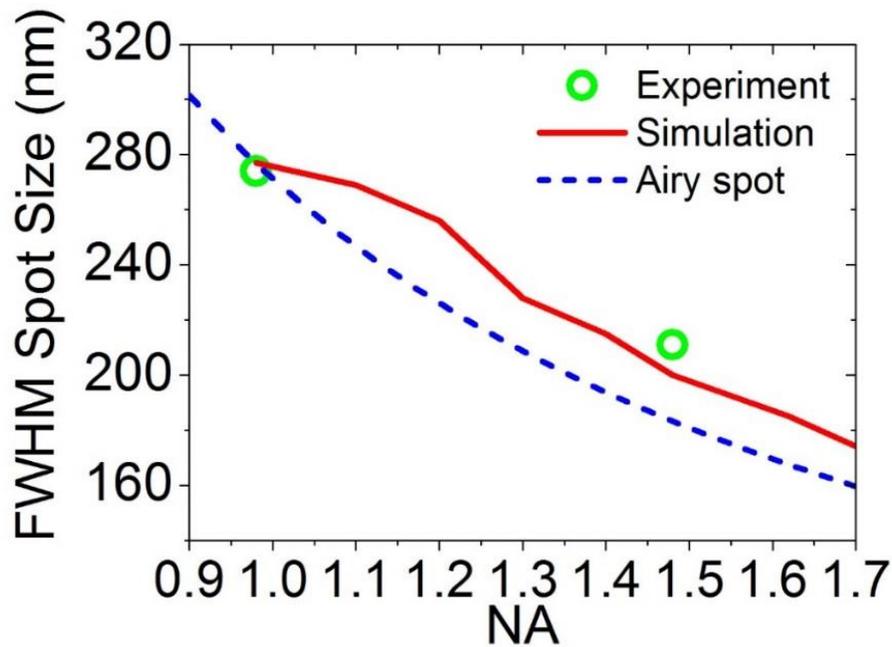

**Figure S2**. Simulated, experimental and theoretical focal spot sizes in terms of NA.